\documentclass[10pt]{article}

\usepackage[margin=1in]{geometry}
\usepackage{lineno}

\usepackage{graphicx}
\usepackage{booktabs}
\usepackage{tabularx}
\usepackage{threeparttable}
\usepackage{ragged2e}
\usepackage{array}
\usepackage{soul}
\usepackage{xcolor}
\usepackage{authblk}
\usepackage{authblk}
\usepackage[numbers,sort&compress]{natbib}
\usepackage{authblk}
\usepackage{graphicx}
\usepackage{booktabs}
\usepackage{tabularx}
\usepackage{threeparttable}
\usepackage{ragged2e}
\usepackage{array, soul}
\usepackage{float}
\usepackage[section]{placeins}
\usepackage{xurl}
\usepackage[hidelinks]{hyperref}
\usepackage{afterpage}

\usepackage{caption}
\DeclareCaptionFont{tinycaption}{\tiny}

\AtBeginDocument{%
  }

\definecolor{lightblue}{rgb}{0.8, 0.92, 0.96}  

\sethlcolor{lightblue}

\newcommand{\highlight}[1]{\relax{#1}}  

\soulregister\cite7
\soulregister\ref7
\soulregister\pageref7
\soulregister\textbf7
\soulregister\emph7

\begin{document}

\title{Plateau That Never Comes: When Efficiency Claims in Datacenters and AI Become Greenwashing}

\author[1]{Harshit Gujral\textsuperscript{*}}
\author[2]{Eshta Bhardwaj\textsuperscript{$\dagger$}}
\author[3]{Dushani Perera\textsuperscript{$\dagger$}}
\author[2]{Christoph Becker\textsuperscript{$o$}}
\author[1]{Steve Easterbrook\textsuperscript{$o$}}

\affil[1]{Department of Computer Science, University of Toronto, Toronto, Ontario, Canada}
\affil[2]{Faculty of Information, University of Toronto, Toronto, Ontario, Canada}
\affil[3]{School of Informatics, University of Edinburgh, Edinburgh, United Kingdom}

\date{}

\maketitle

\begingroup
\renewcommand{\thefootnote}{\fnsymbol{footnote}}
\footnotetext[1]{Corresponding author: \texttt{harshit@cs.toronto.edu}}
\footnotetext[2]{Equally contributed as second authors.}
\renewcommand{\thefootnote}{o}
\footnotetext{Equally contributed as senior authors.}
\endgroup


\begin{abstract}

Datacenter expansion under generative AI is increasingly framed as compatible with sustainability because of efficiency gains, cleaner electricity procurement, and improved facility design. Yet these claims often do not show that absolute electricity, water, material, waste, and community-facing burdens are falling. This Perspective addresses that evidentiary gap. Rather than asking whether efficiency gains are real, we ask when such gains are being enlarged into claims of system-wide sustainability to justify continued expansion. We develop a rebound-informed diagnostic framework for evaluating AI and datacenter sustainability narratives across five tests: metric, boundary, reinvestment, burden shifting, and governance. Applied to major AI industry sustainability reporting, the framework shows that firms largely justify continued expansion through efficiency improvements and clean-energy procurement, rather than by demonstrating reductions in absolute resource use. Applied to plateau claims in the literature, we show that many claims establish local or relative improvements while leaving energy rebound, lifecycle burdens, and enforceable limits unresolved. We argue that these sustainable-growth narratives begin to function as greenwashing when they use efficiency improvements to claim sustainability even as absolute energy, water, material, and public health burdens continue to increase. We conclude by positioning digital sufficiency as a burden-of-proof framework for governance: those advocating further datacenter expansion must show that it reduces, rather than merely redistributes or defers, absolute burdens across the full system.

\end{abstract}


\begin{figure}[H]
\captionsetup{font=footnotesize}
  \centering
   \makebox[\textwidth][c]{%
    \includegraphics[
      width=1.2\textwidth,
      height=0.9\textheight,
      keepaspectratio
    ]{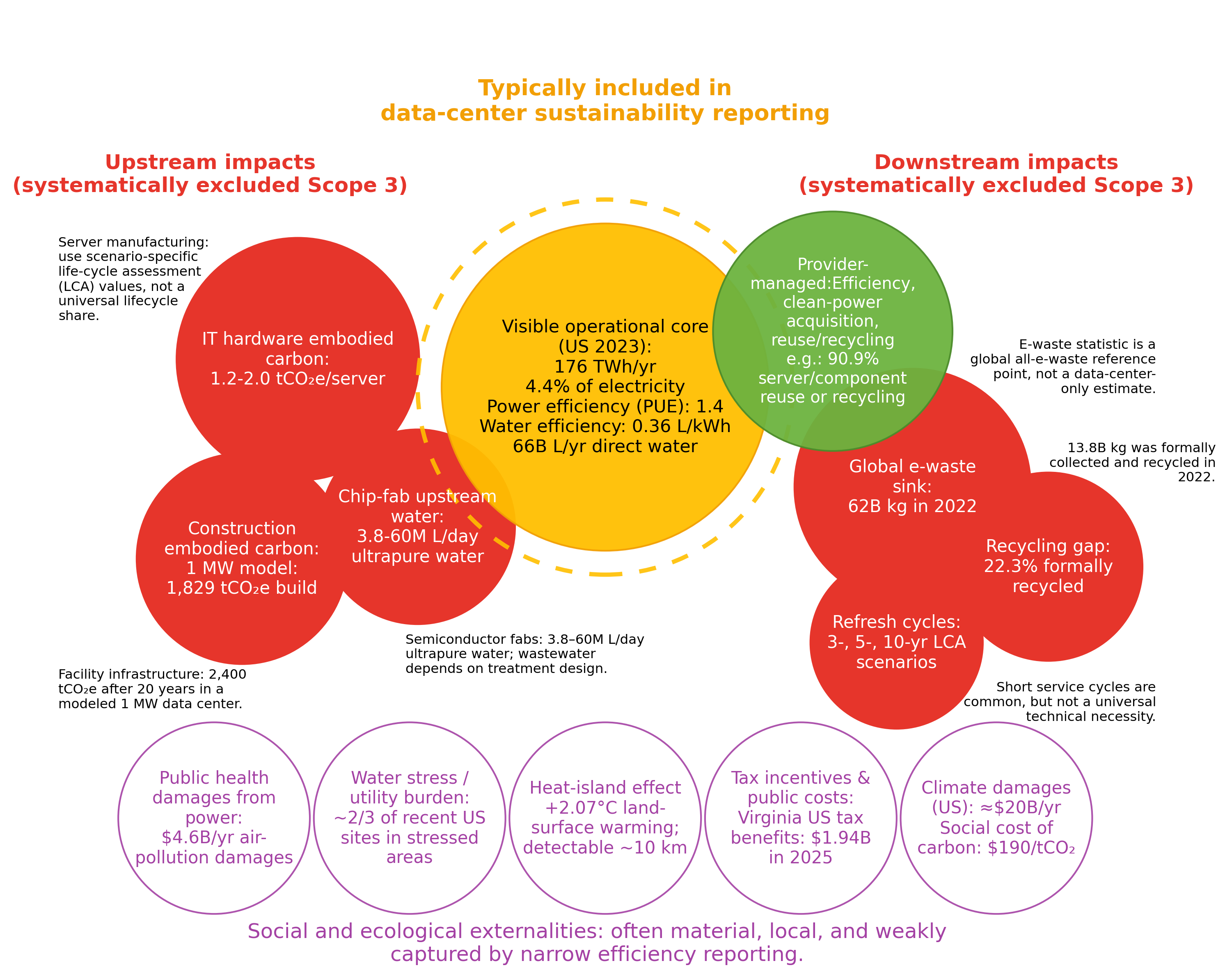}%
  }
  \caption{
The datacenter footprint map makes the boundary problem in datacenter sustainability discourse visible: providers typically optimize and report a relatively narrow set of operational indicators, while many upstream, downstream, and community-facing burdens remain weakly represented, inconsistently quantified, or excluded altogether. Rather than implying comparison across all bubbles, we bring together examples from different scales and evidence types to show the breadth of impacts that surround the operational core of datacenters (yellow bubble)~\cite{shehabi2024data,wadenstein2025life,alissa2025coolclouds,kim2024upw,microsoft2025sustainability,forti2024ewaste,walker2026datacenters,virginia2026tax,marinoni2026dataheat,muller2026datacenters}. The included statistics should be read as representative examples rather than comprehensive. Here, tCO$_2$e denotes tonnes of carbon dioxide equivalent. (The colour palette draws inspiration from Figure 1 in Chapter 4.3 of Greta Thunberg's \emph{The Climate Book}).} \label{fig:dc-footprint-map} 
\end{figure}










\section{Introduction}

Datacenter expansion under generative AI (genAI) is becoming a climate-governance problem because efficiency gains are being used to legitimate growing electricity, water, and public health burdens~\cite{guidi2024environmental}. On current trajectories, there is no clear end in sight for datacenter \highlight{resource demand, as the rise of genAI adds further pressure to expand digital infrastructure.} Sustainability reporting from major technology firms increasingly presents continued datacenter expansion as compatible with sustainability because of efficiency gains, design innovation, and the procurement of renewable energy. Simultaneously, some influential studies argue more strongly that total environmental footprints may plateau or even decline, largely on the assumption that efficiency gains, cleaner electricity, and operational optimization will eventually outpace continued expansion~\cite{iea_data_centres_2023,iea_weo_2024,bhardwaj2025limits,kamiya2025critical}. These claims matter not just as technical forecasts, but as narratives that shape public reassurance, policy imagination, and corporate legitimacy. If the dominant story is that efficiency will eventually bend the curve, then pressure to confront the absolute growth of digital infrastructures and their rising demands on electricity, water, land, and supply chains can be deferred \cite{bhardwaj2025limits,kamiya2025critical}.

\highlight{As firms push LLMs into search, office software, cloud services, developer tools, and consumer platforms, genAI workloads are expanding rapidly despite uneven evidence of user demand and, in some cases, visible user resistance}. A new wave of plateau arguments has, therefore, emerged in the literature: more efficient hardware and model architectures, more capable foundation models, synthetic data, carbon-aware scheduling, expanded renewables, and eventual demand, known as the `headwinds' scenario, will together bend the curve and then shrink it \cite{patterson2022carbon,desislavov2023trends,iea2025energyai,castro2024rethinking}. These arguments sound plausible because they point to real intensity gains, credible technical roadmaps, and familiar decarbonization narratives. But across datacenter and GenAI debates, such narratives share a blind spot: they treat shrinking a subcomponent (per-query energy, power usage effectiveness (PUE), training intensity) as evidence that the whole system will plateau (Fig.~\ref{fig:dc-footprint-map}). The problem is not that these local improvements are not real. The problem is the larger rhetorical move through which shrinking one subcomponent---per-query energy, facility overhead, or training intensity---is enlarged into a claim about the sustainability of the whole system.

Per-unit efficiency gains can be real and important, but they do not by themselves establish system-wide sustainability. By \emph{relative reduction}, we mean lower energy, carbon, or water per unit of computation, query, or workload. By \emph{absolute reduction}, we mean lower total electricity use and lower total water, material, and waste burdens across the relevant system boundary over time (see Fig. \ref{fig:dc-footprint-map}, for the complete system), rather than lower intensity within one stage, facility, or service~\cite{neumayer2010weak}. To evaluate when relative reductions are inflated into claims of \emph{absolute reduction}, we propose five rebound-informed tests: metric, boundary, reinvestment, burden shifting, and governance. These tests distinguish valid efficiency claims from sustainability narratives that provide reputational and policy cover for continued expansion, and that begin to function as greenwashing when absolute impacts remain unbounded.

This work is consequential because datacenters and AI systems are embedded in expansionary business models. Luccioni et al. argue that narrow accounting of their direct operational burdens misses the indirect health effects on communities and the environment, incentive structures, and governance conditions through which efficiency gains can be absorbed into further growth \cite{luccioni2025rebound}. Likewise, efficiency narratives remain institutionally attractive even when practitioners themselves encounter implementation barriers, behavioral complexity, and weak evidence that ``seamless'' optimization delivers system-level sustainability \cite{bremer2025ironies}. The key question  is not whether an efficiency gain occurred. It is whether that gain survives broader system boundaries, avoids reinvestment into scale, does not merely shift burdens elsewhere, and is governed in ways that produce absolute decline.

We apply the proposed tests to two bodies of evidence with different argumentative roles. First, we examine major technology firms’ sustainability reporting to show that corporate narratives do not describe restraint, stabilization, or any end in sight for increasing energy use; they normalize a substitution in which efficiency and innovation stand in for sustainability. Second, we revisit recurring plateau arguments in the datacenter and genAI literature, from historical claims about virtualization, hyperscale efficiency, and low PUE to contemporary claims about model efficiency, synthetic data, carbon-aware scheduling, and demand moderation. The purpose is not to dismiss efficiency improvement work but to establish why cleaner-growth narratives are flawed and why explicit plateau claims do not yet count as evidence of absolute reduction. This reframing also matters for the research community: if indirect and systemic rebound effects are not considered, research and policy can overinvest in efficiency measures that do not deliver their promised sustainability benefits \cite{bremer2023smarthomes}.


\section{Rebound-informed tests for sustainability claims}


This Perspective uses the rebound effect as a diagnostic for how efficiency claims are enlarged into sustainability narratives. Some claims fail because energy savings are reinvested into larger workloads, new features, or wider deployment; others fail because harms are moved, hidden, or postponed through narrow boundaries, burden shifting, or speculative deferral. Rebound and burden-shifting dynamics therefore provide a practical way to test whether local efficiency gains are being translated into absolute reductions~\cite{luccioni2025rebound,bremer2023smarthomes}.



Table~\ref{tab:five-tests} shows five proposed tests to examine the extent of such claims. A claim that fails the metric, boundary, and reinvestment tests remains an efficiency claim even if it refers to a real technical improvement, i.e., it has not yet been shown that the system is becoming sustainable. A claim that also fails the shifting and governance tests, while continuing to market itself as sustainable, net-zero, or plateauing, moves from incomplete sustainability reasoning and begins to function as greenwashing. We use the phrase ``begins to function as greenwashing'' deliberately. The point is not to infer motives. The point is to identify when the social effect of a narrative is to convert local optimization into system-wide reassurance without evidence that total harms are being reduced.

The point of these tests is not only to reject weak claims but to clarify what stronger claims would need to show. A sustainability claim must demonstrate not just a local efficiency gain, but why that gain leads to absolute gains, avoids reinvestment into further growth, does not merely shift burdens elsewhere, and is backed by governance capable of translating local savings into absolute decline.
\afterpage{%
\begin{table*}[t]
\footnotesize
\centering
\begin{threeparttable}
\caption{\highlight{Five rebound-informed tests for evaluating datacenter and AI sustainability claims. The tests ask whether reported improvements concern absolute burdens or only intensity metrics; whether the system boundary includes lifecycle, water, material, and local impacts; whether efficiency gains reduce total demand or are reinvested into more throughput; whether burdens are reduced or shifted elsewhere; and whether binding governance mechanisms constrain expansion, allocation, and hardware turnover. Claims that pass only the first three tests can support efficiency, but not system-level sustainability. Claims that fail the burden-shifting and governance tests while presenting the system as sustainable, net-zero, or plateauing begin to function as greenwashing.}}
\label{tab:five-tests}
\begin{tabularx}{\textwidth}{
>{\RaggedRight\arraybackslash}p{1.3cm}
>{\RaggedRight\arraybackslash}p{2.0cm}
>{\RaggedRight\arraybackslash}p{2.5cm}
>{\RaggedRight\arraybackslash}p{3cm}
>{\RaggedRight\arraybackslash}X}
\toprule
\textbf{Test} & \textbf{Core question} & \textbf{Typical failure} & \textbf{Illustrative example} & \textbf{What would count as stronger evidence} \\
\midrule
Metric 
& Is the claim about intensity or totals? 
& Lower impact per query, workload, or facility overhead is presented as evidence of sustainability. 
& Google reports over 6$\times$ more computing power per unit of electricity than five years earlier,
even as total data-center electricity consumption grew by 27\% in 2024 \cite{google2025environmental}. 
& Evidence that total system burdens---not only per-unit burdens---decline across the relevant period. \\

Boundary 
& What counts as the system? 
& Operational emissions or facility metrics stand in for lifecycle, water, material, and local externalities. 
& AWS foregrounds PUE/WUE and design gains such as up to 46\% lower mechanical energy and 35\% lower concrete embodied carbon \cite{amazon2024sustainability}. 
& Full-system accounting that makes explicit what is included, excluded, and deferred. \\

Reinvestment 
& What happens to the savings? 
& Lower cost, friction, or latency enables more experimentation, adoption, features, and infrastructure build-out. 
& Patterson et al.\ show large training-efficiency gains; the key question is whether such savings are reinvested into more total deployment and inference \cite{patterson2022carbon}. 
& Evidence that efficiency gains are not simply reinvested into more total throughput or demand. \\

Burden Shifting 
& Has the burden fallen, or merely moved? 
& Impacts are displaced across lifecycle stage, geography, time, or accounting category, and narrated as reductions. 
& Microsoft reports Scope 1--2 reductions while total emissions still rose with AI and cloud expansion, highlighting accounting shifts \cite{microsoft2025sustainability}. 
& Demonstration that burdens decline in total rather than being redistributed. \\

Governance 
& \highlight{What binds efficiency gains to lower total demand?}
& \highlight{Future plateau or sustainability is asserted without binding limits on expansion, allocation, or hardware turnover.}
& \highlight{IEA High Efficiency or Headwinds cases lower the curve, but do not consider governance that caps expansion out of scope \cite{iea2025energyai}}. 
& \highlight{Binding energy, water, and material limits; renewable-linked expansion; compute-allocation rules; hardware lifetime and procurement limits; and mandatory lifecycle reporting.} \\
\bottomrule
\end{tabularx}
\end{threeparttable}
\end{table*}
}

\section{Applying tests to AI industry reporting and plateau narratives}

Recent syntheses place 2030 global data-centre electricity demand well above 2020 levels across almost all plausible scenarios, even before water, embodied infrastructure, and supply-chain burdens are fully counted~\cite{kamiya2025critical,bhardwaj2025limits}. We therefore apply the five tests to two influential sites where local efficiency gains are often scaled into system-level reassurance: AI industry reporting (Table~\ref{tab:AI industry-merged}) and prevalent plateau narratives in research (Table~\ref{tab:plateau-families}).

\subsection{AI industry reporting as a diagnostic tool}

The clearest evidence that there is no plateau in sight comes from the firms themselves. This is not because these firms promise a plateau and fail to deliver it; it is because their own reporting does not claim one at all. Across Google, AWS, Microsoft, Meta, and Equinix, sustainability communications present a consistent pattern: rising demand is acknowledged, efficiency and cleaner procurement are foregrounded, and absolute growth is treated as compatible with sustainability \cite{google2025environmental,amazon2024sustainability,microsoft2025sustainability,meta2024sustainability,equinix2025sustainabilityreport,equinix2025press}. Table~\ref{tab:AI industry-merged} shows what this substitution looks like in practice and where each firm's reporting falls short under the five tests.

Google, for instance, reports year-over-year growth in datacenters' electricity consumption driven by AI adoption while simultaneously emphasizing hardware efficiency and low fleet-wide PUE \cite{google2025environmental}. Meta is even more direct, pairing large year-over-year growth in electricity use with a continuing net-zero-operational narrative built on procurement and accounting \cite{meta2024sustainability}. Microsoft couples its 2030 carbon and carbon-free-electricity goals to continued cloud and AI expansion rather than to any commitment to limit compute or demand \cite{microsoft2025sustainability}. AWS stresses new cooling architectures, custom chips, and better energy and water metrics as it builds out additional AI-ready capacity \cite{amazon2024sustainability}. Equinix similarly highlights PUE improvement, renewable coverage, and net-zero ambition while continuing to expand colocation and AI-oriented infrastructure \cite{equinix2025sustainabilityreport,equinix2025press}. What is being promised across all five cases is not modest infrastructure, fewer workloads, or lower total throughput, but cleaner growth.

Under the five-test framework, three failures recur:
\begin{itemize}
    \item \textbf{Metric substitution:} improving PUE, reducing energy per workload, or matching purchased electricity with a cleaner supply says something about a component of the system, but does not establish that total burdens are falling.
    
    \item \textbf{Drawing the boundary too narrowly:} the operational core is made highly legible while broader upstream and downstream impacts remain hidden or undercounted.
    
    \item \textbf{Governance displacement:} net-zero commitments, renewable matching, and supply-chain engagement are presented as if equivalent to binding mechanisms that would cap or reduce total throughput.
\end{itemize}

Yet, the disclosures in Table~\ref{tab:AI industry-merged} do not describe restraint, stabilization, or any credible cap on absolute demand. They describe continued expansion, enabled by efficiency measures to sustain further growth, rather than constrained by legally binding obligations to account for its environmental and public health impacts~\cite{guidi2024environmental}.

\subsection{The plateau story keeps returning}

Some influential, highly cited studies make a different and stronger move. Here, the efficiency measures are used not merely to legitimate cleaner growth, but to argue that total sectoral electricity demand can remain roughly flat or even decline. Before the genAI boom, several influential assessments argued that datacenter electricity demand could remain roughly flat even as workloads grew, provided that server efficiency, virtualization, hyperscale migration, and low PUE continued to improve \cite{koomey2011growth,shehabi2016united,iea2017digitalisation,masanet2020recalibrating}. These studies did not all make identical claims, and they were often more careful than the public story that followed them~\cite{sciencedaily2016plateau,mcmahon2020wired}. 

Nevertheless, they helped entrench a reassuring idea: digitalization might \emph{decouple} from energy growth quickly enough that the sector's expanding role need not be treated as a major sustainability problem. \highlight{What made that reassurance plausible was the broader idea that continued growth can be separated from the rising carbon and energy burdens of expansion. In this view, datacenters and AI could keep scaling while greenhouse-gas emissions fall quickly and durably enough to contain their overall footprint. That broader claim has been challenged repeatedly: observed improvements are too partial, too local, too temporary, or too slow to support a system-wide sustainability conclusion, and promised future reductions remain weak without governance that limits total scale~\cite{vaden2020decoupling,parrique2019evidence,haberl2020systematic}.}

The new wave of genAI plateau narratives updates this reassurance rather than replacing it. Unlike the corporate reports above, these arguments do attempt to say something about total trajectories. The object of optimization has shifted — from server utilization and facility overhead to accelerators, model architectures, and inference pathways — but the underlying blind spot remains the same: a measured improvement in a component of the system is taken as evidence that the overall trajectory is under control. Table~\ref{tab:plateau-families} organizes the current claims into three families. Efficiency optimism holds that better accelerators, lower-precision computation, algorithmic improvements, longer-lived foundation models, and synthetic data will keep energy per task falling \cite{patterson2022carbon,desislavov2023trends,castro2024rethinking}. Operational burden shifting holds that carbon-aware scheduling, cleaner siting, and edge or on-device inference will redistribute workloads toward lower-impact locations or times \cite{patterson2022carbon,castro2024rethinking}. Demand moderation by assumption holds that after an initial boom, adoption will slow, best practices will diffuse, and the curve will bend \cite{castro2024rethinking,kamiya2025critical}, \highlight{often through scenario logic that lowers the curve without yet demonstrating the governance that would make such moderation real \cite{castro2024rethinking,iea2025energyai}.} What changes across these families is the site of optimization. What does not change is the leap from local efficiency gain or \emph{relative reduction} to system-level stabilization or \emph{absolute reduction}.

The subsections that follow examine how these narratives are inflated using the five tests: first, through metric and boundary substitution; then, through reinvestment; next, through burden shifting; and finally, through speculative deferral and absent governance.

\subsection{Metric and boundary substitution}

The most common failure is metric substitution: a reduction in the energy or emissions intensity of one operation is treated as evidence that the total system is becoming sustainable. Boundary narrowing then reinforces this move by counting only the operational core while hiding embodied carbon, water, e-waste, supply chains, and local externalities.


Recent studies have discussed the increased efficiencies achieved by AI hardware and infrastructure, including decreased average PUE, server energy intensity, number of servers per workload, and storage drive energy use \cite{masanet2020recalibrating}. Reports also discuss best practices that reduce emissions from training ML models \cite{masanet2020recalibrating}. The question is not whether such gains are real, but what they actually establish about total system sustainability.



This can be seen with PUE: due to historic gains in PUE, big datacenters currently operate near low-PUE (approximately 1.1--1.2), suggesting the potential to hold datacenter facility energy flat \cite{shehabi2016united}. However, the overall energy usage of datacenters continues to grow despite these PUE improvements. IT load growth, including accelerated servers, negates savings and instead drives net energy usage. PUE is valuable for managing overhead and identifying inefficient operations, but it does not tell us whether total electricity use is falling, whether IT load itself is growing, what embodied impacts attach to new hardware, or how local water and grid burdens are changing. When PUE improvement is allowed to stand in for sustainability, the metric test and the boundary test both fail simultaneously.

The same is true when corporate reporting foregrounds facility and procurement improvements while backgrounding supply chains, embodied carbon, and local infrastructure effects. Such claims may still be useful as statements about efficiency, but they do not, on their own, establish sustainability.

\subsection{Reinvestment and rebound}

\highlight{If the metric and boundary tests ask what is being measured and what is being left out, the next question is what the resulting savings enable.} Rebound effects in AI cannot be reduced to technical performance alone, and technical efficiency does not guarantee net environmental reductions because deployment is shaped by business incentives, governance, and broader social norms \cite{luccioni2025rebound}. Once an efficiency gain lowers the cost, friction, or latency of computation, the key question becomes what those savings enable. In the AI context, these dynamics reflect the political economy of deployment: when the dominant market incentive is to expand usage, increase product integration, and intensify platform dependence, efficiency gains become the fuel for further growth rather than the basis for absolute reduction.

LLM training requires substantial energy, water, and emissions, and recent plateau narratives suggest that this burden may moderate as fewer large-scale training runs are needed and models are instead reused, distilled, or fine-tuned \cite{Luccioni_envprimer_2024,Sevilla_2022,patterson2022carbon,castro2024rethinking}. But a lower training burden does not yet establish a lower total footprint. Lower training frequency can coexist with growing burdens of imprint costs, ranging from repeated fine-tuning, evaluation, retrieval augmentation, safety systems, synthetic-data generation, to orchestration, and, above all, inference at scale. The relevant question is therefore not whether one category of cost has fallen, but what those savings enable at system scale \cite{Luccioni_powerhungry_2024}.

A related claim is that synthetic data and longer-lived models reduce the need for repeated retraining of the largest models~\cite{patterson2022carbon}. Even if they do, that does not yet establish a lower total footprint. Costs can shift into data generation, filtering, storage, evaluation, and inference, especially when synthetic data is produced \highlight{by large source models, which are used to train other models}~\cite{Shao_2024}. The issue is not whether synthetic data yields local savings, but whether those savings hold once the full system is counted.

The same logic applies to product integration more broadly. Once AI capabilities become cheaper to serve, they are embedded across search, office suites, cloud services, developer tools, devices, and enterprise platforms. Each integration can be defended locally as a marginal addition made more feasible by efficiency gains. But a paper or report that documents lower energy per task while leaving the fate of those savings unexamined fails the reinvestment test.

\subsection{Burden shifting rather than reducing}

Not all weak sustainability claims are strict rebound cases. Some fail because burdens are shifted rather than reduced: across lifecycle stages, geography, time, or accounting categories. In genAI literature, this gap manifests as shifting the burden among different phases of the lifecycle: inference-side optimization may lower per-query energy use, but once models are widely deployed, burdens can move from training to ongoing model inference, imprint, supporting services, and downstream use. \highlight{The reverse can also occur: additional upstream compute may be used to reduce inference-time cost, shifting burdens back into pre-training, post-training imprints, or distillation rather than removing them \cite{deepseekv3,arora2026training}.}


Lifecycle burden shifting is now common in AI debates more broadly. Training often receives the most public attention, yet once models become widely deployed, inference, supporting services, and repeated downstream use can dominate the system's ongoing footprint. For instance, integrating AI search into a conventional search engine---and making it the default---can raise the baseline burden of ordinary querying by shifting more interactions onto inference-intensive systems. A narrative that points to stabilizing training intensity while ignoring the rise of inference is therefore not a narrative of reduction---it is a narrative of burden moving to another stage of the lifecycle. Likewise, edge or on-device inference can reduce some network and datacenter burdens while increasing device-side computation, battery impacts, replacement cycles, and embedded hardware demand. Whether such shifts are less damaging to the environment is an empirical question; \highlight{such reductions cannot be assumed} from lower burden in one location alone.

Efforts to introduce computational efficiencies at the time of inference have focused on reducing per-query energy use. This includes using mixture of experts (MoE) models that ``selectively activate only the parts of the model most pertinent to solving the query in question, thereby saving on computation and hence energy costs while preserving model performance'' \citep[p.~43]{iea2025energyai}. It can also include ``operational optimizations such as [adjusting] batch sizes, key-value cache management, and attention management.'' \citep[p.~43-44]{iea2025energyai}. These efforts towards cutting energy usage per query would be plausible if the demand for AI systems remained constant. However, AI applications continue to be embedded across applications \cite{Beignon_Thibault_Maudet_2025} and their usage continues to grow (observed by the millions of LLM users) \cite{Bellan_2025}, indicating that inference-side efficiency will not outrun usage.

Temporal and spatial burden shifting create a similar problem. Carbon-aware scheduling can move work toward hours or regions with lower operational emissions intensity, and cleaner siting can relocate facilities toward grids with more renewable supply or better ambient conditions for cooling. These practices may indeed reduce the emissions associated with specific runs. But burden shifting is not a reduction unless the total burden falls after the broader consequences are counted. Cleaner windows can become new windows for expansion, and new sites can transfer pressure to regions with different water regimes, land conflicts, or grid constraints. Under the five-test framework, the move from ``cleaner here or now'' to ``less overall'' requires evidence, not assumption.

Accounting shifts are subtler but equally consequential. Corporate net-zero narratives may improve a metric within Scope 2 or a narrowly defined operational boundary while leaving Scope 3 burdens, local infrastructure impacts, and hardware turnover largely out of frame. The environmental harm has not disappeared; it has become harder to see from the angle at which the system is being reported. This is one reason it is argued that a narrow focus on direct emissions misrepresents AI’s climate footprint \cite{luccioni2025rebound}.

\subsection{Speculative deferral and absent governance}


The weakest plateau claims defer the problem into the future by relying on anticipated breakthroughs, expected slowdowns, or hoped-for market corrections rather than present pathways to absolute reduction. Scenario variants in which high efficiency, operational flexibility, or demand ``headwinds'' lower future trajectories may be useful~\cite{iea2025energyai}, but a possibility is not yet a governed pathway.


The most hopeful plateau claims look toward ``the future of computing.'' This includes horizon technologies such as photonic, neuromorphic, or quantum approaches \cite{iea2025energyai}. But it also includes the belief that advances in AI across science and technology domains will enable the development of a growth infrastructure that does not rely on removing or working around current constraints of the energy market, or the current paradigms of pursuing algorithmic performance gains \cite{iea2025energyai}. \highlight{What underwrites that optimism is the broader idea that continued growth can be separated from the rising energy use and greenhouse-gas emissions that expansion would otherwise bring. But wider critiques of this claim have shown that observed reductions are often too partial, too local, too temporary, or too slow to justify that level of reassurance~\cite{vaden2020decoupling,haberl2020systematic,parrique2019evidence}.}

This is where the governance test becomes decisive. A forecast that assumes future best-practice adoption, slower demand growth, or a cleaner electricity supply may be internally coherent. But if the mechanisms that would produce that outcome are unspecified, voluntary, or politically unlikely, the forecast does not establish sustainability. It establishes only that, under favorable assumptions, the curve could be lower than it otherwise would have been. That is not the same thing as showing that the system is being governed toward absolute decline.

Even in the scenario where such ``drop-in'' solutions are realized, datacenter facility turnover and software stacks lag by years, thus continuing the resource usage. More efficient cooling systems, novel chips, advanced memory hierarchies, optical interconnects, or future breakthroughs in computing do not escape the earlier tests. They still need to survive reinvestment, burden shifting, and governance. A more efficient future datacenter can still be part of an expanding, more materially intensive, and more spatially concentrated digital regime. More concerning is that such optimism can delay present governance around resource use and demand while current social and ecological impacts continue to accumulate.

Bremer et al.'s account of techno-optimism is useful here. In their study, professionals held strong beliefs in the future efficacy of digital efficiency interventions even while recounting substantial barriers and contradictory evidence from practice \cite{bremer2025ironies}. The analogy is not exact, but the pattern is recognizable: the more speculative the plateau claim becomes, the more it relies on the seductiveness of optimization rather than on demonstrated institutional capacity to constrain growth. Under the five-test framework, the problem is not uncertainty alone, but the displacement of present governance by faith in future correction.

If speculative futures cannot substitute for present governance, the next question is what kind of governance could translate local efficiency gains into an absolute reduction. This is where digital sufficiency becomes necessary.

\section{Digital sufficiency as the governance condition for absolute reduction}

\highlight{Datacenter expansion is placing additional burdens on health and well-being in communities, often in places already facing disproportionate environmental stress (Fig. ~\ref{fig:dc-footprint-map}). 
The AI industry's desire for expansion can only be sustainable if these total burdens fall. This means not only less energy per query, lower PUE, or cleaner electricity per workload, but lower absolute impacts across the system. This is the core decoupling challenge: can datacenter expansion continue without a corresponding rise in environmental impact? The existing literature thoroughly debunks the idea that we can reduce absolute carbon emissions while continuing to scale the infrastructure~\cite{parrique2019evidence,haberl2020systematic,vaden2020decoupling}. At the current pace of expansion, sustainability would require carbon emissions from the system to fall at the same time as digital activity grows; yet relative improvements in efficiency often coexist with rising total impacts~\cite{parrique2019evidence,haberl2020systematic,vaden2020decoupling}. 

We argue that those advocating further datacenter expansion should bear the burden of proving that such expansion will be environmentally sustainable. This requires governance capable of evaluating and enforcing absolute limits, rather than merely tracking intensity improvements: whether claimed reductions are absolute or only per-unit, whole-system, or only facility-level, and whether they reduce current burdens rather than defer them to future technologies or shift them to grids, watersheds, chip supply chains, and e-waste destinations. The concept of \emph{sufficiency}, which has emerged in the sustainability literature, describes the condition of having just enough of something rather than too little or too much. Santarius et al. define \emph{digital sufficiency} as ``any strategy aimed at directly or indirectly decreasing the absolute level of resource and energy demand from the production or application of ICT'', encompassing hardware, software, user, and economic sufficiency~\cite{santarius2023digital}. Digital sufficiency can provide a framework for governance. It asks how much genAI infrastructure is justified, which uses of genAI compute should be prioritized, and what constraints are needed to prevent genAI efficiency gains from being reinvested into further scale~\cite{santarius2023digital,freitag2021real, knowles2025computingresponsibility}. In this section, we develop digital sufficiency as a burden-of-proof framework for expansion, specifying what sufficiency-informed governance would require across hardware and datacenter infrastructure, software and model design, and economic incentives (Section~\ref{sub:hardware_software_economic}) and user practices (Section~\ref{sub:user}).}

\subsection{What governance would actually mean}
\label{sub:hardware_software_economic}

\highlight{If digital sufficiency is a burden-of-proof framework for expansion, then governance must identify where expansion is produced and where limits can be enforced. For the AI industry,} this can be translated into three governance dimensions: (i) governing infrastructure growth, hardware procurement, workload allocation, (ii) software design, and (iii) the economic incentives that turn efficiency gains into new demand.

\highlight{\emph{The first dimension} concerns infrastructure growth and facility operation. Current datacenter design prioritizes maximum performance, redundancy, and constant availability; sufficiency-oriented governance would instead ask what computing capacity is justified under local energy, water, grid, and emissions constraints, and what existing capacity exceeds those limits or serves uses that cannot be publicly justified.} This can be understood through three governance areas:

\begin{enumerate}

    \item First, \emph{operational provisioning limits} would ensure that datacenter growth does not add to community stress through their public-health impacts from fossil-fuel-based energy generation and water stress. In addition, new expansion approvals should be tied to additional renewable and grid capacity, caps on total electricity use, and strategies such as pre-cooling or thermal storage to shift flexible workloads toward periods when renewable electricity is abundant and reduce activity when it is scarce~\cite{Madon2020}. These measures should count as sufficiency only when they reduce absolute burdens or prevent new demand from exceeding local energy, water, health, and infrastructure limits.

    \item \highlight{Second, \emph{demand-side allocation of compute} would govern how constrained capacity is distributed across workloads, models, products, and features. If compute, electricity, or water availability is limited, facilities should not treat all computing workloads as equally necessary. Public-interest, reliability-critical, accessibility, scientific, and climate-relevant workloads could be prioritized over discretionary, speculative, or engagement-maximizing uses during scarcity periods.}
    
    \item Third, \emph{material and procurement limits} would ensure that hardware expansion does not simply transfer the burdens of digital growth onto communities affected by extraction, manufacturing, energy use, water use, and e-waste. Rather than assuming continual refresh cycles and unconstrained accelerator procurement, firms should have to justify new hardware against existing utilization, lifecycle emissions, critical mineral demand, repair and reuse options, and expected e-waste~\cite{Hilty2015}. The governance point is not only to make hardware cleaner, but to challenge the assumption that frequent replacement is necessary for progress.
\end{enumerate}

\emph{The second dimension} concerns software and model design. Datacenter demand is not produced by hardware alone; it is also produced by software defaults, model architectures, product features, and interface choices that determine how often computation is invoked. In genAI, capability gains have often been pursued through scaling---larger models, larger datasets, and more frequent deployment---which increases both training and inference burdens~\cite{bhardwaj2025limits}. A sufficiency-oriented approach would therefore ask different design questions: when are smaller or more specialized models enough; when should defaults avoid continuous or automatic integration and invocation of AI models; and how can interfaces support lower-impact choices without offloading responsibility entirely onto users?~\cite{gujral2025design}. 

Governance would have to make these design choices visible and contestable. This requires disclosure not only of training costs, but also of inference demand, model size, deployment frequency, automatic invocation, feature-level compute requirements, and the conditions under which smaller models or non-genAI alternatives would be sufficient. Such disclosure would make it harder to treat algorithmic efficiency as sustainability while total use keeps expanding. It could be paired with procurement standards, default-setting rules, and incentives for genAI systems that meet task requirements with lower computational demand. Without such constraints, gains in hardware and algorithmic efficiency are likely to be absorbed into larger models, more features, and more frequent inference.

\highlight{\emph{The third dimension} concerns economic governance of datacenters. The rapid expansion of genAI infrastructure is driven by business models organized around scale, utilization, market dominance, and continuous product integration. Governance cannot treat private investment in genAI infrastructure as a neutral market outcome. Following credit-guidance and public-purpose provisioning arguments, capital flows into genAI should be regulated and justified according to democratically defined public needs, including community wellbeing~\cite{hickel2024creditguidance}.} This investment problem is also an antitrust problem because when a few firms control compute, cloud infrastructure, chips, foundation models, and deployment channels, they also control the conditions under which genAI expands: who gets access and what uses are prioritized.

Antimonopoly approaches to governing the AI industry point to a broader toolkit than after-the-fact antitrust enforcement: structural separation, nondiscrimination and open-access rules, interoperability standards, scrutiny of acquisitions and exclusive cloud--model partnerships, public options, and cooperative governance~\cite{narechania2024antimonopoly,korinek2025concentrating}. These tools matter for digital sufficiency because control over compute, cloud, chips, foundation models, and deployment channels determines who can build genAI systems, what kinds of systems are economically viable, and whether efficiency gains are translated into restraint or further scale. If a few vertically integrated firms control these bottlenecks, smaller and lower-impact alternatives will struggle to compete against models built around scale.

\subsection{Creating user agency through digital sufficiency}
\label{sub:user}
When it comes to governing user-side consumption of AI, digital sufficiency is not about personal austerity. It is about managing the absolute burden of the system and giving users a choice to act in accordance with their environmental values. 

Many digital services are intentionally designed to maximize user engagement and data consumption, contributing to rising demand for datacenters. Governance should therefore require digital-sufficiency mechanisms to be embedded directly into the user interfaces of digital services and platforms. This includes building awareness of the carbon footprint of digital activity, providing transparency about both system-level and user-facing footprints, and offering sufficiency-based controls such as slower sync, storage caps, scheduled downtime, longer retrieval times, and higher-latency or lower-compute modes where appropriate. Such mechanisms can give users greater voice and agency to act in accordance with their environmental values without making them solely responsible for systemic emissions. Existing work on sustainable datacenters and user preferences suggests that many users are more open to sufficiency-oriented options than dominant design assumptions imply~\cite{gujral2025design}. When efficiency is interpreted through a rebound-informed framework and tied to governance mechanisms that confront total scale, it can become part of a credible sustainability pathway. The practical and research challenge is not whether computation can be made more efficient, but whether efficiency gains are governed in ways that produce absolute reductions at the system scale. These three dimensions illustrate how digital sufficiency could transform the trajectory of datacenters and AI infrastructures. The question facing policymakers, researchers, and industry actors is therefore not whether computation can become more efficient, but whether society is willing to define how much computation is enough.

\section*{Acknowledgements}
We thank Ma\"el Madon, Dylan Van Bramer, and Sara Hooker for their participation in early brainstorming sessions that helped shape the initial direction of this project.

\section*{Competing interests}
The authors declare no known conflict of interests. 

\section*{Funding statement}
HG and SME acknowledge the support from the Natural Sciences and Engineering Research Council of Canada (NSERC, RGPIN-2019-07042).


\nolinenumbers
\small
\bibliographystyle{naturemag}
\bibliography{sample-base}

\appendix
\renewcommand{\thetable}{\Alph{section}.\arabic{table}}
\setcounter{table}{0}
\renewcommand{\thetable}{\Alph{section}.\arabic{table}}
\setcounter{table}{0}

\newpage
\section{Supplementary tables}
\begin{table}[!ht]
\footnotesize
\centering
\begin{threeparttable}
\caption{AI industry sustainability reporting frames continued expansion as manageable, not bounded, and cleaner growth. Columns show the dominant framing, how each of the five tests fails, and what the framing leaves undemonstrated.}
\label{tab:AI industry-merged}
\begin{tabularx}{\linewidth}{>{\RaggedRight\arraybackslash}p{1.0cm} >{\RaggedRight\arraybackslash}X >{\RaggedRight\arraybackslash}p{1.5cm} >{\RaggedRight\arraybackslash}X}
\toprule
\textbf{Firm} & \textbf{Representative sustainability framing} & \textbf{Failed tests} & \textbf{What the framing leaves undemonstrated} \\
\midrule
Google & AI-related electricity demand is rising, but hardware, model, and operational efficiency are presented as tempering energy and emissions growth \cite{google2025environmental}. & Metric, boundary, governance & No demonstrated pathway from improved intensity metrics to lower total datacenter demand; no commitment to restrain AI-driven growth or to account fully for wider embodied and supply-chain burdens. \\
\addlinespace
AWS & New datacenter components, custom chips, and cooling systems are presented as reducing mechanical energy use and improving PUE/WUE while AWS builds new AI-ready facilities \cite{amazon2024sustainability}. & Metric, boundary, governance & Expansion is presumed and mitigated, not limited; better facility metrics do not demonstrate a lower total footprint across electricity, water, and hardware supply chains. \\
\addlinespace
Microsoft & Carbon, water, and 24/7 carbon-free electricity goals are maintained alongside rapid cloud and AI growth \cite{microsoft2025sustainability}. & Boundary, burden shifting, governance & Operational and procurement goals are emphasized without showing how continued growth is reconciled with absolute reductions across full lifecycle and supply-chain boundaries. \\
\addlinespace
Meta & Rising datacenter electricity use tied to AI deployment is paired with continued net-zero-operational claims built around renewables, PUE, and accounting mechanisms \cite{meta2024sustainability}. & Metric, burden shifting, governance & A narrative of greener operations coexist with expanding electricity demand and infrastructure; procurement and accounting do not establish lower total electricity, material, or water burdens. \\
\addlinespace
Equinix & Lower PUE, higher renewable coverage, and net-zero-by-2040 targets accompany continued colocation and AI-infrastructure growth \cite{equinix2025sustainabilityreport,equinix2025press}. & Metric, boundary, governance & Improved intensity metrics are not evidence of lower system demand; no mechanism is shown that would convert efficiency improvements into bounded total throughput. \\
\bottomrule
\end{tabularx}
\end{threeparttable}
\end{table}

\begin{table}[!ht]
\scriptsize
\centering
\begin{threeparttable}
\caption{Recurring plateau arguments in the literature and scenario reports: \highlight{the hidden assumption each claim relies on}, the tests they fail, and what stronger evidence would require.}
\label{tab:plateau-families}
\begin{tabularx}{\linewidth}{>{\RaggedRight\arraybackslash}p{2.0cm} >{\RaggedRight\arraybackslash}X >{\RaggedRight\arraybackslash}p{1.8cm} >{\RaggedRight\arraybackslash}X}
\toprule
\textbf{Claim family} & \textbf{Representative claim and hidden move} & \textbf{Failed tests} & \textbf{What stronger evidence would require} \\
\midrule
Historical datacenter plateau & Efficiency gains in servers, virtualization, and facility overhead are treated as evidence that total sector electricity can stay roughly flat, generalizing from a subset of improving metrics to sector-wide trajectories \cite{koomey2011growth,shehabi2016united,masanet2020recalibrating}. & Metric, boundary, governance & Demonstrated decline in total system footprint at sector scale, with full accounting for rebound from new workloads and for embodied burdens. \\
\addlinespace
Training plateau & Large reductions in training emissions from best practices are rhetorically extended to AI's overall trajectory, while costs shift into fine-tuning, evaluation, synthetic-data pipelines, and inference-heavy deployment \cite{patterson2022carbon}. & Reinvestment, burden shifting, governance & Evidence that training gains are not offset by scaling, fine-tuning, evaluation, synthetic-data generation, and inference growth. \\
\addlinespace
Inference efficiency optimism & Lower energy per inference or better hardware utilization is taken as evidence of a lower total AI burden, without accounting for rising served demand, uptime, and feature expansion \cite{desislavov2023trends,google2025environmental}. & Metric, reinvestment & Evidence that total served demand, uptime, and feature expansion do not absorb efficiency gains. \\
\addlinespace
Synthetic data and longer-lived models & Reduced need for \highlight{repeated retraining of the largest models (known as frontier retraining)} is assumed to lower overall footprint, while costs shift into data generation, filtering, storage, evaluation, and downstream inference and are left unaccounted \cite{patterson2022carbon,castro2024rethinking}. & Reinvestment, burden shifting, boundary & Lifecycle accounting across training, synthetic-data generation, evaluation, and deployment, showing whether reduced retraining is outweighed by expanded service use. \\
\addlinespace
Temporal and spatial burden shifting & Cleaner timing, cleaner siting, or edge and on-device pathways are treated as reductions rather than redistributions of burden, without accounting for water, land, device-side, and embodied impacts \cite{patterson2022carbon,iea2025energyai}. & Burden shifting, boundary, governance & Evidence that total burdens fall after full accounting, plus demand controls ensuring that cleaner windows do not simply enable more total work. \\
\addlinespace
Headwinds and horizon technologies & Speculative future slowdowns, breakthrough technologies, or near-universal best-practice diffusion are allowed to substitute for present governance, deferring the burden of proof into an assumed future \cite{castro2024rethinking,iea2025energyai}. & Governance, reinvestment & Demonstrated near-term mechanisms for constraining throughput or footprint, rather than optimism about future correction. \\
\bottomrule
\end{tabularx}
\end{threeparttable}
\end{table}

\end{document}